\documentclass{Interspeech2024}

\interspeechcameraready

\title{Analyzing Multimodal Features of Spontaneous Voice Assistant Commands for Mild Cognitive Impairment Detection}

\name[affiliation={1}]{Nana}{Lin}
\name[affiliation={1}]{Youxiang}{Zhu}
\name[affiliation={1}]{Xiaohui}{Liang}
\name[affiliation={2}]{John A.}{Batsis}
\name[affiliation={2}]{Caroline}{Summerour}

\address{
  $^1$University of Massachusetts Boston, MA, USA\\
  $^2$University of North Carolina, Chapel Hill, NC, USA}
\email{nana.lin002@umb.edu, youxiang.zhu001@umb.edu, xiaohui.liang@umb.edu, john.batsis@unc.edu, caroline\_summerour@med.unc.edu}

\keywords{speech analysis, machine learning, multimodal model, mild cognitive impairment}

\usepackage{caption}
\usepackage{multirow}

\begin{document}

\maketitle

\begin{abstract}
    
    Mild cognitive impairment (MCI) is a major public health concern due to its high risk of progressing to dementia. This study investigates the potential of detecting MCI with spontaneous voice assistant (VA) commands from 35 older adults in a controlled setting. Specifically, a command-generation task is designed with pre-defined intents for participants to freely generate commands that are more associated with cognitive ability than read commands. We develop MCI classification and regression models with audio, textual, intent, and multimodal fusion features. We find the command-generation task outperforms the command-reading task with an average classification accuracy of 82\%, achieved by leveraging multimodal fusion features. In addition, generated commands correlate more strongly with memory and attention subdomains than read commands. Our results confirm the effectiveness of the command-generation task and imply the promise of using longitudinal in-home commands for MCI detection. 

\end{abstract}

\section{Introduction}
Mild cognitive impairment (MCI) presents a significant public health concern because of its substantial risk of progressing to dementia~\cite{whoDementia}. Around half of MCI patients develop dementia within 3 years and an annual progression rate of $6\%$ to $15\%$~\cite{breton2019cognitive}. 
Early detection of MCI is crucial for preventing or delaying the emergence of dementia~\cite{rasmussen2019alzheimer}. Speech-based dementia detection is promising as a low-cost early-stage screening method~\cite{fristed2021evaluation,vigo2022speech,liang2022evaluating,yang2022deep}.
Researchers have explored various approaches for detecting MCI and other cognitive impairments from speech, such as automated phone task~\cite{marshall2015harvard}, picture description task~\cite{luz2021detecting, zhu2023evaluating}, and tablet-based automatic assessment~\cite{yamada2021tablet}.

Previous works mainly focused on the common dataset, including Pitt Corpus and the Cookie Theft Picture description task~\cite{mueller2018connected, luz2020alzheimer, luz2021detecting}. Despite three decades of research, this task faces data limitations due to the expensive data collection process~\cite{zhu2023evaluating}. In recent years, an AI-driven Voice Assistant (VA) can be integrated into a wide range of Internet of Things (IoT) devices, such as smart speakers and smartphones. They can engage older adults in conversations, which makes it a practical and convenient tool for collecting speech data and enabling health-related speech-based research~\cite{trajkova2020alexa, purao2021use, ruggiano2021chatbots, liang2022evaluating, upadhyay2023studying}. 
Researchers~\cite{kurtz2023early} evaluated VA capabilities in detecting MCI through a command-reading task with a focus on analyzing audio data embeddings. Both MCI and Healthy Controls (HC) participants demonstrated highly similar command outputs when reading from identical instruction sheets of VA commands.
We consider the speech dataset collected from such a command-reading task does not require the participant to use much of their memory and attention ability, known as cognitive load~\cite{sweller2011cognitive}. Cognitive load has been proven as an effective metric to be used to assess a patient's mental condition~\cite{yamada2021tablet}. We envision that a task that gives older adults more freedom in the command generation process represents a more realistic VA scenario in their homes and can enhance MCI detection.

In this paper, we design a command-generation task through the VA system to enable participants to freely generate commands according to an intent, which may result in subtle changes between MCI and HC. Specifically, the command-generation task provides participants with a set of intent keywords to assist them in generating voice commands. The commands can be formed differently but meant to accomplish the same intent.  
We propose new intent features and aim to study and compare the effectiveness of the intent features, and fusion of multimodal features extracted from the VA commands and investigate whether the spontaneous VA commands from the command-generation task can achieve enhanced performance in MCI detection. The contributions of our paper are as follows:

First, we propose a command-generation task through VA, which is associated with a higher cognitive load, resulting in a better performance in MCI detection than the command-reading task. The average classification accuracy of all features in multi-modality achieved higher results, increased from 73\% to 77\% for all classifiers, and increased from 78\% to 82\% using the Random Forest (RF) classifier. 

Second, we study the correlation between the VA commands and six subdomains of the Montreal Cognitive Assessment (MoCA)~\cite{nasreddine2005montreal}. VA commands from the command-generation task exhibit stronger correlations across all subdomains compared to the command-reading task, particularly within memory and attention subdomains.
The results show that the command-generation task demands increased attention and short-term memory retention.

Third, we compare the performance of intent, audio, textual, and multimodal fusion features on the command-generation task and find that audio and fusion (intent + audio) features achieved better classification accuracy (mean: 75\% and 74\%, best: 88\% and 91\%). The results confirm that the intent and audio features are more sensitive to the MCI detection than the textual features.

\begin{figure*}[]
    \centerline{\includegraphics[width=0.95\textwidth]{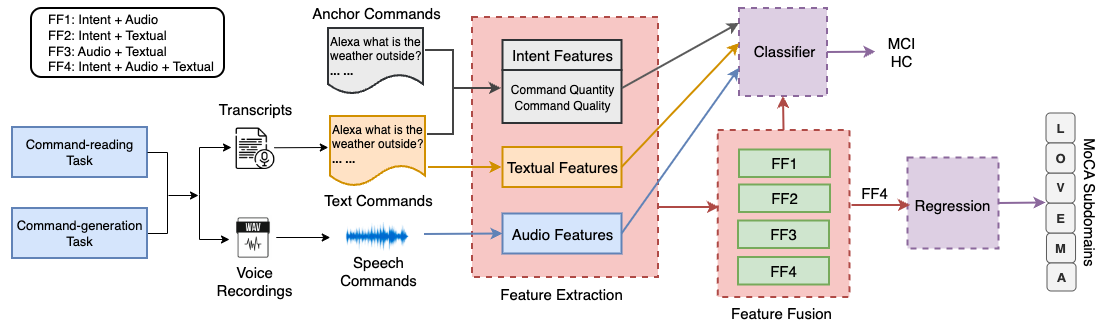}}
\caption{MCI detection model using multimodal features extracted from voice assistant commands (L: Language, O: Orientation, V: Visuospatial, E: Executive function, M: Memory, A: Attention).}
\vspace{-0.2cm}
\label{fig2}
\end{figure*}

\section{Voice Assistant Dataset}
To study the effectiveness of the VA commands for MCI detection, we conducted both structured (\textbf{Command-reading}) and semi-structured (\textbf{Command-generation}) tasks. The audio recordings and the
corresponding Automatic Speech Recognition (ASR) transcripts were collected from older adults.

\subsection{Participant Recruitment \& Data Collection}
We recruited 35 older adults aged 65 and older for a longitudinal study consisting of seven sessions from December 2021 to January 2024, with each session every three months. Among the participants, there were 16 females and 19 males, with an average age of 72 years old. 
We assessed their MoCA score and subdomain scores~\cite{wood2020montreal, hobson2015montreal}, and used them as labels in the classification and regression models. The MoCA consists of six subdomains, with 30 scores in total: Memory $\leq$ 15, Executive Function $\leq$ 13, Attention $\leq$ 18, Language $\leq$ 6, Visuospatial $\leq$ 7, Orientation $\leq$ 6.
Based on the MoCA threshold, MoCA score $\geq 26$ indicates HC, otherwise MCI~\cite{nasreddine2005montreal, kurtz2023early}. 15 participants were diagnosed with MCI, while 20 were deemed healthy in their first sessions. 
Throughout the study, a participant could not complete sessions 6 and 7 due to health problems. We have finished 243 interviews with 98 MCI and 145 HC labels.

\subsection{Perform Voice Assistant Commands}
In every session, the participants were asked to interact with Alexa devices for two speech tasks under the research assistant’s (RA) guidance.
The examples of VA commands and intent keywords designed for the two tasks are listed in Table~\ref{tab:task2}. We selected common commands used by older adults and defined 6 categories~\cite{liang2022evaluating, alexaElderly, alexaGuide}: information, entertainment, productivity, shopping, communication, and smart home.

\textbf{Command-reading} task: Participants read 34 commands to the Alexa where 30 common commands include ``What is the weather outside," ``Remind me to start the laundry tomorrow at 2 pm" and 4 short commands include ``yes" and ``pause."

\textbf{Command-generation} task: Aligning with the command-reading task, we provided 34 intent keywords, such as ``check weather", ``play music", and ``add a reminder". The participants need to generate the voice commands based on these intents. The commands can be formed differently but meant to accomplish the same intent. 

\begin{table}[th]
    \caption{{Examples of commands and intents of VA tasks}}
    \label{tab:task2}
    \centering
    \Huge
    \renewcommand{\arraystretch}{1.5}
    \resizebox{0.48\textwidth}{!}{%
        \begin{tabular}{c|c|c}
        \hline
        \textbf{Category} & \textbf{Commands} & \textbf{Intents} \\ 
        \hline
        Information & ``What is the weather outside?" & Check weather \\
        \hline
        Entertainment & ``Play classical music." & Play music\\
        \hline
        Productivity & ``Remind me to start the laundry tomorrow at 2 PM." & Add reminder, Laundry\\
        \hline
        Shopping & ``Add oranges and grapes to my shopping list." & Add shopping list, Fruits \\
        \hline
        Communication & ``Call (603)660-2203."  & Make a phone call \\
        \hline
        Smart Home & ``Turn on the bedroom light" & Turn on light\\	
        \hline
        \end{tabular}
    }
    \vspace{-0.4cm}
\end{table}

\section{Methodology}

In this research, we implemented two multimodal classification and regression models for MCI detection. Figure \ref{fig2} illustrates the architecture of the classification model between MCI and HC, and the regression model between the VA commands and the six MoCA subdomains, using multimodal features in the command-reading and command-generation tasks.

\subsection{Feature Extraction}
\begin{figure}[]
    \centerline{\includegraphics[width=\linewidth]{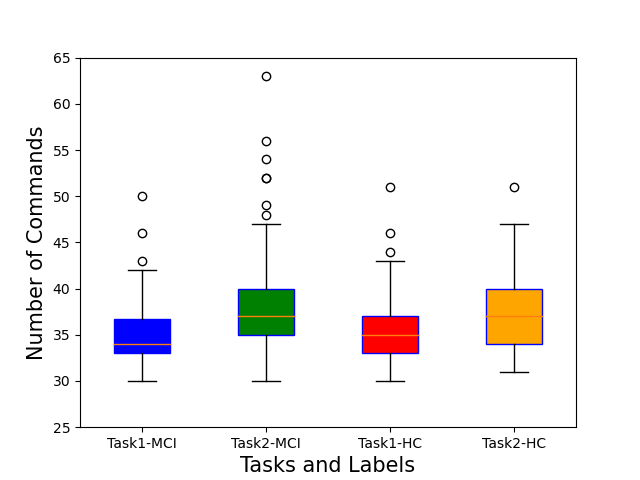}}
\caption{Box plot of command count comparison in command-reading (Task1) and command-generation (Task2) tasks.}
\vspace{-0.45cm}
\label{fig1}
\end{figure}

\textbf{Intent Features}. 
We observed that the number of commands from the command-generation task is larger than the command-reading task, especially for the MCI, as shown in Figure~\ref{fig1}.
The command-generation task, with more outliers in MCI than HC, has a larger difference in the interquartile range between MCI and HC than the command-reading task.
Based on this observation, we encoded the commands with Sentence Transformer \textbf{all-mpnet-base-v2} model~\cite{huggingface2021} and calculated the cosine similarity between the anchor commands and collected commands. The anchor commands are the provided 34 commands.
We created the intent features by concatenating the following quantity features and quality features.
The \textbf{quantity feature}  is the count of the collected commands aligning with specific anchor commands. 
Given that all participants can finish the whole process, a smaller number of commands indicates that they can perform the task faster and more concentrated. 
The \textbf{quality feature} is the average cosine similarity between collected commands and specific anchor commands.
A larger average similarity indicates that generated commands are closer to the provided commands or intents, implying the different cognitive abilities related to command reading and generation process. 
The formulas for calculating the quantity ($QTY$) and quality ($QLT$) features are as follows. $\texttt{sim}(i, j)$ is the cosine similarity between the $i_{th}$ anchor command and the $j_{th}$ collected command, where $i \in (0, n)$, $j \in (0, m)$, $n$ is the number of anchor commands, and $m$ is the number of collected commands. $c_i$ is the count of collected commands most relevant to the $i_{th}$ anchor commands.

{\footnotesize
\[
QTY(i) = \sum_{j=0}^{m} \begin{cases}
1 & \text{if } \texttt{sim}(i, j) = \max_{k=0}^{n} \texttt{sim}(i, j) \\
0 & \text{otherwise}
\end{cases}
\]

\[
QLT(i) = \frac{1}{c_i} \sum_{j=0}^{m} \begin{cases}
\texttt{sim}(i, j) & \text{if } \texttt{sim}(i, j) = \max_{k=0}^{n} \texttt{sim}(i, j) \\
0 & \text{otherwise}
\end{cases}
\]
}

\textbf{Textual Features.}
We extracted the textual features from the ASR transcripts using \textbf{Bidirectional Encoder Representations from Transformers (BERT)} model~\cite{devlin2018bert}. It involves utilizing the pre-trained language model to extract dense representations of input text. Textual features derived from BERT embeddings encode semantic, syntactic, and contextual information within high-dimensional vectors. We consider that more freedom in the command generation process compared to the command reading process will require more cognitive load from participants, which improves the effectiveness of textual features analysis.

\textbf{Audio Features.}
After sampling rate normalization to 16kHz, we extracted the audio features from the speech data using the \textbf{Wav2vec2} processor and the \textbf{Hidden-Unit BERT (HuBERT)} model~\cite{hsu2021hubert}.
The fine-tuned HuBERT utilizes self-supervised speech representation learning through offline clustering to align target labels for a BERT-like prediction loss.
Audio features derived from the HuBERT embeddings explore the hierarchical representations of speech data, including spectral patterns, phonetic nuances, and semantic elements. 

\textbf{Fusion Features in Multi-modality.}
We integrated audio, textual, and intent features to leverage the combination of different modalities within a model, including FF1 (Intent + Audio), FF2 (Intent + Textual), FF3 (Audio + Textual), FF4 (Intent + Audio + Textual). 
It encapsulates the process of merging features from multi-modalities into a unified representation.
By incorporating audio features from spoken language, textual features from transcriptions, and intent features from the quantity and quality of the commands, our model possesses the power of multimodal features fusion. 
This integration allows for a comprehensive analysis and understanding of VA commands, enabling the model to interpret and process information from multiple dimensions simultaneously in MCI detection.

\subsection{MCI Detection Model}
\textbf{Classification.} We implement the classification model to explore the effectiveness of spontaneous VA commands in MCI detection.
We first extract three distinct feature sets from the VA dataset using sentence transformer, speech, and language processing models, encompassing intent, audio, and textual features. Then we use four fusion features that combine the previous three features in multi-modality. 
All the feature sets are then integrated into the downstream classifier to create a binary classification task to predict either MCI or HC. Each feature set was classified using four different models: Decision Tree (DT), Linear Support Vector Machine (SVM)~\cite{scikit-learn}, K Nearest Neighbors (KNN), and RF~\cite{breiman2001random}.
We report the mean and best results including accuracy, precision, recall, and F1-score.

\textbf{Regression.}
To observe the correlation between the VA commands and the MoCA subdomain scores, we choose Linear Ridge Regression (LRR), DT, and SVM models as the downstream regression models with the FF4 (Intent + audio + textual) fusion features of the VA dataset, using MoCA subdomain scores as labels. Due to the different magnitudes of the predictions and the comparison between different datasets, we introduced the Relative Root Mean Square Error (RRMSE)~\cite{despotovic2016evaluation, hernandez2000application} as a normalized measure. The average Mean Absolute Error (MAE), Root-Mean-Square Error (RMSE), and RRMSE of MoCA subdomain scores are reported.

\section{Experiments}
\subsection{Implementation Details}
\textbf{Data Preprocessing.} 
After segmenting the audio and transcript files of the 243 records on a command-by-command basis for both tasks, we deleted the error, unmatched and non-participant commands. Error commands include VA unrecognized contents, for example, ``audio could not be understood". Unmatched commands have either audio data or transcript data missing. Non-participant commands are what RA demonstrated to the participants or commands said by other non-participants during the interview.

We collected $m$ commands per participant, which are between 30 and 65. We calculated the intent features in the shape of the provided 34 commands and 34 intents. We extracted the audio embedding features (shape: $m \times 1024$) and the textual embedding features (shape: $m \times 768$) from $m$ commands.
To align the shape of features, we calculated the average embeddings of all commands and concatenated them sequentially for all the features in multi-modality. 

\textbf{Experimental Settings.}
All experiments were performed on multiple NVIDIA Tesla A100 GPUs.
Due to the relatively small dataset, nested 10-fold cross-validation~\cite {brownless2020nested} was used with the 10 rounds in the outer loop for 100 trials. 
All metrics are obtained in model-specific trials.

\begin{table}[ht]
    \centering
    \caption{Classification results of command-reading (Task1) and command-generation (Task2) tasks. Better results are in bold.}
     \resizebox{0.48\textwidth}{!}{%
    \begin{tabular}{c|cc|cc|cc|cc}
    \hline
        \multirow{2}{*}{Model} & \multicolumn{2}{c|}{Accuracy} &  \multicolumn{2}{c|}{Precise} & \multicolumn{2}{c|}{Recall} & \multicolumn{2}{c}{F1} \\ 
        ~ & Task1 & Task2 & Task1 & Task2 & Task1 & Task2 & Task1 & Task2 \\ \hline
        \rule{0pt}{10pt}
        DT & 0.74 & \textbf{0.81} & 0.77 & \textbf{0.86} & \textbf{0.94} & 0.90 & 0.76 & \textbf{0.82} \\
        SVM & 0.71 & \textbf{0.75} & 0.75 & \textbf{0.78} & 0.95 & \textbf{0.96} & 0.81 & \textbf{0.83} \\ 
        KNN & 0.70 & 0.70 & \textbf{0.78} & 0.77 & 0.80 & \textbf{0.82} & 0.75 & 0.75 \\
        RF & 0.78 & \textbf{0.82} & 0.82 & \textbf{0.83} & 0.93 & \textbf{0.94} & 0.84 & \textbf{0.86} \\ \hline
        \rule{0pt}{10pt}
        Avg. & 0.73 & \textbf{0.77} & 0.78 & \textbf{0.81} & 0.91 & 0.91 & 0.79 & \textbf{0.82} \\ \hline
    \end{tabular}}
    \vspace{-0.4cm}
    \label{classification}
\end{table}

\begin{table*}[ht]
    \centering
    \Huge
    \caption{Correlation between VA commands and MoCA subdomains in command-reading and command-generation tasks. Better RRMSE results in two tasks are in bold.}
    \resizebox{0.98\textwidth}{!}{%
    \begin{tabular}{c|ccc|ccc|ccc|ccc|ccc|ccc}
    \hline
        \multirow{3}{*}{Subdomain} & \multicolumn{9}{c|}{Command-reading Task} & \multicolumn{9}{c}{Command-generation Task} \\ \cline{2-19}
        \rule{0pt}{10pt}
        ~ & \multicolumn{3}{c|}{DT} & \multicolumn{3}{c|}{SVR} & \multicolumn{3}{c|}{LRR} & \multicolumn{3}{c|}{DT} & \multicolumn{3}{c|}{SVR} & \multicolumn{3}{c}{LRR}\\
        ~ & MAE & RMSE & RRMSE & MAE & RMSE & RRMSE & MAE & RMSE & RRMSE & MAE & RMSE & RRMSE & MAE & RMSE & RRMSE & MAE & RMSE & RRMSE \\ \hline
        \rule{0pt}{10pt}
        memory & 1.79 & 1.82 & 3.90\% & 1.31 & 1.70 & 2.84\% & 1.30 & 1.65 & 2.62\% & 1.59 & 2.03 & \textbf{3.45\%} & 1.31 & 1.69 & \textbf{2.83\%} & 1.27 & 1.68 & \textbf{2.59\%} \\ 
        executive function & 1.42 & 2.23 & 3.70\% & 1.07 & 1.83 & 2.71\% & 1.13 & 1.66 & 2.60\% & 1.52 & 2.25 & \textbf{3.64\%} & 1.07 & 1.83 & \textbf{2.70\%} & 1.10 & 1.66 & \textbf{2.52\%} \\
        attention & 1.90 & 2.19 & \textbf{3.15\%} & 1.59 & 1.73 & 2.71\% & 1.55 & 1.57 & 2.53\% & 1.94 & 2.29 & 3.30\% & 1.58 & 1.73 & \textbf{2.70\%} & 1.47 & 1.61 & \textbf{2.45\%} \\
        language & 0.70 & 1.18 & \textbf{3.81\%} & 0.60 & 0.86 & 3.31\% & 0.60 & 0.72 & \textbf{2.74\%} & 0.72 & 0.98 & 3.86\% & 0.60 & 0.85 & 3.31\% & 0.62 & 0.72 & 2.82\% \\
        visuospatial & 0.94 & 1.43 & 4.59\% & 0.70 & 1.11 & 3.38\% & 0.76 & 1.08 & 3.39\% & 0.90 & 1.14 & \textbf{4.42\%} & 0.70 & 1.12 & 3.38\% & 0.73 & 1.12 & \textbf{3.28\%} \\
        orientation & 0.59 & 0.75 & \textbf{3.27\%} & 0.44 & 0.50 & 2.50\% & 0.50 & 0.45 & \textbf{2.29\%} & 0.58 & 0.47 & 3.29\% & 0.44 & 0.50 & 2.50\% & 0.51 & 0.43 & 2.32\%\\ \hline
        \rule{0pt}{10pt}
        MoCA & 2.56 & 3.45 & 2.80\% & 2.10 & 2.30 & 2.11\% & 1.93 & 2.42 & 1.92\% & 2.28 & 3.34 & \textbf{2.33\%} & 2.09 & 2.30 & \textbf{2.10\%} & 1.85 & 2.45 & \textbf{1.83\%} \\ \hline
    \end{tabular}
    }
    \label{regression}
\end{table*}

\begin{table*}[ht]
    \centering
    \caption{Comparison of classification results across multimodal features using RF classifier (MCI vs. HC). Accuracy $\geq 80\%$ and F1-score $\geq 80\%$ are underlined. Better accuracy results in two tasks are in bold.}
    \resizebox{0.98\textwidth}{!}{%
    \begin{tabular}{c|cccccccc|cccccccc}
    \hline
        \rule{0pt}{10pt}
        \multirow{3}{*}{Feature} & \multicolumn{8}{c|}{Command-reading Task} & \multicolumn{8}{c}{Command-generation Task} \\ \cline{2-17}
        ~ & \multicolumn{2}{c}{Accuracy} & \multicolumn{2}{c}{Precision} & \multicolumn{2}{c}{Recall} & \multicolumn{2}{c|}{F1} & \multicolumn{2}{c}{Accuracy} & \multicolumn{2}{c}{Precision} & \multicolumn{2}{c}{Recall} & \multicolumn{2}{c}{F1} \\ 
        ~ & Mean & Best & Mean & Best & Mean & Best & Mean & Best & Mean & Best & Mean & Best & Mean & Best & Mean & Best \\ \hline
        Intent & 0.50 & 0.65 & 0.58 & 0.67 & 0.64 & 0.83 & 0.60 & 0.71 & 0.55 & \textbf{0.74} & 0.61 & 0.77 & 0.67 & 0.92 & 0.63 & \underline{0.80} \\ 
        Audio & 0.72 & \underline{0.83} & 0.74 & 0.88 & 0.86 & 0.88 & 0.78 & \underline{0.88} & 0.75 & \underline{\textbf{0.88}} & 0.75 & 0.93 & 0.89 & 0.88 & \underline{0.81} & \underline{0.90} \\ 
        Textual & 0.62 & 0.74 & 0.65 & 0.76 & 0.81 & 0.92 & 0.72 & 0.78 & 0.62 & \underline{\textbf{0.83}} & 0.66 & 0.80 & 0.81 & 0.92 & 0.72 & \underline{0.86} \\ 
        FF1 & 0.68 & 0.79 & 0.71 & 0.87 & 0.82 & 0.94 & 0.75 & \underline{0.86} & 0.74 & \underline{\textbf{0.91}} & 0.73 & 0.94 & 0.91 & 1.00 & \underline{0.80} & \underline{0.94} \\ 
        FF2 & 0.65 & \underline{\textbf{0.90}} & 0.68 & 0.87 & 0.83 & 1.00 & 0.74 & \underline{0.93} & 0.61 & 0.70 & 0.64 & 0.71 & 0.81 & 0.92 & 0.71 & 0.78 \\ 
        FF3 & 0.69 & 0.75 & 0.69 & 0.80 & 0.88 & 0.92 & 0.77 & \underline{0.82} & 0.73 & \underline{\textbf{0.87}} & 0.73 & 0.87 & 0.90 & 0.94 & \underline{0.80} & \underline{0.90} \\ 
        FF4 & 0.65 & 0.78 & 0.68 & 0.85 & 0.82 & 1.00 & 0.74 & \underline{0.85} & 0.71 & \underline{\textbf{0.83}} & 0.71 & 0.82 & 0.89 & 1.00 & 0.79 & \underline{0.87} \\ \hline
        Avg. & 0.65	& 0.78	& 0.68	& 0.82	& 0.82	& 0.93	& 0.73	& \underline{0.84} & 0.67	& \underline{\textbf{0.82}} & 	0.69 &	0.83 & 0.84 & 0.94& 0.75 & \underline{0.86} \\ \hline
    \end{tabular}
    }
     \vspace{-0.4cm}
    \label{classification2}
\end{table*}

\subsection{Command-reading vs. Command-generation Tasks}
\textbf{Classification Evaluation.}
Table \ref{classification} shows the average classification results of all seven feature sets for both tasks employing four classifiers (DT, SVM, KNN, and RF). The command-generation task outperforms the command-reading task with an increased average accuracy of all models from 73\% to 77\% and the RF classifier achieved the best accuracy of 82\%. 
The classification results of using the RF classifier leveraging multimodal features are shown in Table \ref{classification2}.
All the metrics of intent, audio, textual, FF1, FF3, and FF4 features from the command-generation task are better than the command-reading task.
In particular, the intent features increased by 9\% from 65\% to 74\%.
The average accuracy of all seven features increases from 78\% to 82\%.
This demonstrates that the command-generation task involves a higher cognitive load and is more effective than the command-reading task in MCI detection. The command-reading task simplifies the real-world VA interaction and has limited cognitive loads. However, participants in the command-generation task interact with the VA by generating spontaneous commands, which involve various cognitive abilities. 

\textbf{Cognitive Subdomains Correlation.}
Table \ref{regression} shows the correlation results between the VA commands and MoCA subdomains from command-reading and command-generation tasks with fusion features FF4 (Intent + Audio + Textual) in three regression models (LRR, DT, and SVM).
We observe that the RRMSE from all three regression models in the command-generation task is decreased compared with the command-reading task with the minimum RRMSE of the MoCA score of 1.83\% in the LRR model. It shows that the generated commands have higher correlations with the MoCA score than the read commands. Such difference is observed in subdomains of memory, attention, executive function, and visuospatial.
We consider that the relatively stronger correlations with the memory (2.45\%) and attention (2.59\%) subdomains confirm that older adults performing the command-generation task with a higher cognitive load require increased attention and short-term memory retention.

\subsection{Features Evaluation}
Table \ref{classification2} compares the classification results of all multimodal features for both tasks.
The audio and FF2 features achieve better accuracy of 83\% and 90\% than the other features in the command-reading task.
The command-generation task performs better with the audio, textual, FF1, FF3, and FF4 features compared to other features. In particular, the accuracy of audio, FF1, and FF3 features reached 88\%, 91\%, and 87\%, respectively, as the top-performing features.
Our results demonstrate the potential of the multimodal fusion features for MCI detection using the VA commands and prove the effective design of the command-generation task. Audio features have higher performance than textual features in MCI and HC classification for both tasks.
The textual feature achieved a mean accuracy of 62\%, with a maximum of 83\% for the command-generation task. 
Especially, the mean accuracy of audio features is increased around  $13\%$ compared with the textual features from $62\%$ to $75\%$ in the command-generation task.

\section{Conclusion and Future Work} 
In this paper, we design and evaluate a VA command-generation task for MCI detection. Results show the command-generation task achieved the best accuracy of $91\%$ by using the FF1 features in the classification model. The average accuracy of the seven features in the command-generation outperforms the command-reading task from 78\% to 82\% using the RF classifier, demonstrating the VA task with a higher cognitive load has better performance. Furthermore, we assessed the correlation between the VA commands and cognitive subdomains. 
Our study reveals that the memory and attention subdomains exhibit stronger correlations with the generated commands. The results confirm the effectiveness of our command-generation task to detect MCI.
In addition, we evaluate the multimodal features of VA commands and show the audio and FF1 features have the best performances among all feature sets for the command-generation task.

We envision that in-home VA commands can be passively collected from older adults in the long term and using them for cognitive health monitoring is significant. One associated challenge is that older adults may repeatedly use familiar commands in their daily routines, which limits the effectiveness of in-home data.
Therefore, using in-home data for cognitive health monitoring needs further investigation.
The traditional VA functionality is limited by the constraints of AI capabilities.
Large generative models like chatGPT and Gemini offer the potential for advanced dialogue capabilities within VA. Future works include the exploration of artificial general intelligence, as well as the investigation of in-home commands for MCI detection.

\section{Acknowledgements}


This research is funded by the US National Institutes of Health National Institute on Aging R01AG067416. We thank Professor David Kotz and Professor Brian MacWhinney for the feedback about this work.

\bibliographystyle{IEEEtran}
\bibliography{reference}

\begin{thebibliography}{10}
\providecommand{\url}[1]{#1}
\csname url@samestyle\endcsname
\providecommand{\newblock}{\relax}
\providecommand{\bibinfo}[2]{#2}
\providecommand{\BIBentrySTDinterwordspacing}{\spaceskip=0pt\relax}
\providecommand{\BIBentryALTinterwordstretchfactor}{4}
\providecommand{\BIBentryALTinterwordspacing}{\spaceskip=\fontdimen2\font plus
\BIBentryALTinterwordstretchfactor\fontdimen3\font minus \fontdimen4\font\relax}
\providecommand{\BIBforeignlanguage}[2]{{%
\expandafter\ifx\csname l@#1\endcsname\relax
\typeout{** WARNING: IEEEtran.bst: No hyphenation pattern has been}%
\typeout{** loaded for the language `#1'. Using the pattern for}%
\typeout{** the default language instead.}%
\else
\language=\csname l@#1\endcsname
\fi
#2}}
\providecommand{\BIBdecl}{\relax}
\BIBdecl

\bibitem{whoDementia}
\BIBentryALTinterwordspacing
{{World Health Organization}}. {``{D}ementia"}. 2023. [Online]. Available: \url{https://www.who.int/news-room/fact-sheets/detail/dementia}
\BIBentrySTDinterwordspacing

\bibitem{breton2019cognitive}
A.~Breton, D.~Casey, and N.~A. Arnaoutoglou, ``Cognitive tests for the detection of mild cognitive impairment (mci), the prodromal stage of dementia: Meta-analysis of diagnostic accuracy studies,'' \emph{International journal of geriatric psychiatry}, vol.~34, no.~2, pp. 233--242, 2019.

\bibitem{rasmussen2019alzheimer}
J.~Rasmussen and H.~Langerman, ``Alzheimer’s disease--why we need early diagnosis,'' \emph{Degenerative neurological and neuromuscular disease}, pp. 123--130, 2019.

\bibitem{fristed2021evaluation}
E.~Fristed, C.~Skirrow, M.~Meszaros, R.~Lenain, U.~Meepegama, S.~Cappa, D.~Aarsland, and J.~Weston, ``Evaluation of a speech-based ai system for early detection of alzheimer’s disease remotely via smartphones,'' \emph{medRxiv}, pp. 2021--10, 2021.

\bibitem{vigo2022speech}
I.~Vigo, L.~Coelho, and S.~Reis, ``Speech-and language-based classification of alzheimer’s disease: A systematic review,'' \emph{Bioengineering}, vol.~9, no.~1, p.~27, 2022.

\bibitem{liang2022evaluating}
X.~Liang, J.~A. Batsis, Y.~Zhu, T.~M. Driesse, R.~M. Roth, D.~Kotz, and B.~MacWhinney, ``Evaluating voice-assistant commands for dementia detection,'' \emph{Computer Speech \& Language}, vol.~72, p. 101297, 2022.

\bibitem{yang2022deep}
Q.~Yang, X.~Li, X.~Ding, F.~Xu, and Z.~Ling, ``Deep learning-based speech analysis for alzheimer’s disease detection: a literature review,'' \emph{Alzheimer's Research \& Therapy}, vol.~14, no.~1, pp. 1--16, 2022.

\bibitem{marshall2015harvard}
G.~A. Marshall, M.~Dekhtyar, J.~M. Bruno, K.~Jethwani, R.~E. Amariglio, K.~A. Johnson, R.~A. Sperling, and D.~M. Rentz, ``The harvard automated phone task: new performance-based activities of daily living tests for early alzheimer’s disease,'' \emph{The journal of prevention of Alzheimer's disease}, vol.~2, no.~4, p. 242, 2015.

\bibitem{luz2021detecting}
S.~Luz, F.~Haider, S.~de~la Fuente, D.~Fromm, and MacWhinney, ``Detecting cognitive decline using speech only: The adresso challenge,'' \emph{arXiv preprint arXiv:2104.09356}, 2021.

\bibitem{zhu2023evaluating}
Y.~Zhu, N.~Lin, X.~Liang, J.~A. Batsis, R.~M. Roth, and B.~MacWhinney, ``Evaluating picture description speech for dementia detection using image-text alignment,'' \emph{arXiv preprint arXiv:2308.07933}, 2023.

\bibitem{yamada2021tablet}
Y.~Yamada, K.~Shinkawa, M.~Kobayashi, M.~Nishimura, M.~Nemoto, E.~Tsukada, M.~Ota, K.~Nemoto, and T.~Arai, ``Tablet-based automatic assessment for early detection of alzheimer's disease using speech responses to daily life questions,'' \emph{Frontiers in Digital Health}, vol.~3, p.~30, 2021.

\bibitem{mueller2018connected}
K.~D. Mueller, B.~Hermann, J.~Mecollari, and L.~S. Turkstra, ``Connected speech and language in mild cognitive impairment and alzheimer’s disease: A review of picture description tasks,'' \emph{Journal of clinical and experimental neuropsychology}, vol.~40, no.~9, pp. 917--939, 2018.

\bibitem{luz2020alzheimer}
S.~Luz, F.~Haider, S.~de~la Fuente, D.~Fromm, and B.~MacWhinney, ``Alzheimer's dementia recognition through spontaneous speech: The adress challenge,'' \emph{arXiv preprint arXiv:2004.06833}, 2020.

\bibitem{trajkova2020alexa}
M.~Trajkova and A.~Martin-Hammond, ``" alexa is a toy": exploring older adults' reasons for using, limiting, and abandoning echo,'' in \emph{Proceedings of the 2020 CHI conference on human factors in computing systems}, 2020, pp. 1--13.

\bibitem{purao2021use}
S.~Purao, H.~Hao, and C.~Meng, ``The use of smart home speakers by the elderly: exploratory analyses and potential for big data,'' \emph{Big Data Research}, vol.~25, p. 100224, 2021.

\bibitem{ruggiano2021chatbots}
N.~Ruggiano, E.~L. Brown, L.~Roberts, C.~V. Framil~Suarez, Y.~Luo, Z.~Hao, and V.~Hristidis, ``Chatbots to support people with dementia and their caregivers: systematic review of functions and quality,'' \emph{Journal of medical Internet research}, vol.~23, no.~6, p. e25006, 2021.

\bibitem{upadhyay2023studying}
P.~Upadhyay, S.~Heung, S.~Azenkot, and R.~N. Brewer, ``Studying exploration \& long-term use of voice assistants by older adults,'' in \emph{Proceedings of the 2023 CHI Conference on Human Factors in Computing Systems}, 2023, pp. 1--11.

\bibitem{kurtz2023early}
E.~Kurtz, Y.~Zhu, T.~Driesse, B.~Tran, J.~A. Batsis, R.~M. Roth, and X.~Liang, ``Early detection of cognitive decline using voice assistant commands,'' in \emph{IEEE International Conference on Acoustics, Speech and Signal Processing (ICASSP)}, 2023, pp. 1--5.

\bibitem{sweller2011cognitive}
J.~Sweller, ``Cognitive load theory,'' in \emph{Psychology of learning and motivation}.\hskip 1em plus 0.5em minus 0.4em\relax Elsevier, 2011, vol.~55, pp. 37--76.

\bibitem{nasreddine2005montreal}
Z.~S. Nasreddine, N.~A. Phillips, V.~B{\'e}dirian, S.~Charbonneau, V.~Whitehead, I.~Collin, J.~L. Cummings, and H.~Chertkow, ``The montreal cognitive assessment, moca: a brief screening tool for mild cognitive impairment,'' \emph{Journal of the American Geriatrics Society}, vol.~53, no.~4, pp. 695--699, 2005.

\bibitem{wood2020montreal}
J.~L. Wood, S.~Weintraub, C.~Coventry, J.~Xu, H.~Zhang, E.~Rogalski, M.-M. Mesulam, and T.~Gefen, ``Montreal cognitive assessment (moca) performance and domain-specific index scores in amnestic versus aphasic dementia,'' \emph{Journal of the International Neuropsychological Society}, vol.~26, no.~9, pp. 927--931, 2020.

\bibitem{hobson2015montreal}
J.~Hobson, ``The montreal cognitive assessment (moca),'' \emph{Occupational Medicine}, vol.~65, no.~9, pp. 764--765, 2015.

\bibitem{alexaElderly}
\BIBentryALTinterwordspacing
{``Amazon Echo and Alexa for the Elderly"}. 2021. [Online]. Available: \url{https://www.techenhancedlife.com/explorers/amazon-echo-and-alexa-elderly}
\BIBentrySTDinterwordspacing

\bibitem{alexaGuide}
\BIBentryALTinterwordspacing
{``Alexa Guide for Seniors: 14 Ways Older Adults Can Use Amazon Echo Devices"}. 2021. [Online]. Available: \url{https://www.vivint.com/resources/article/alexa-guide-for-seniors}
\BIBentrySTDinterwordspacing

\bibitem{huggingface2021}
HuggingFace, ``All-mpnet-base-v2 (v2.0.0) [pre-trained model for multilingual text classification],'' Retrieved from \url{https://huggingface.co/all-mpnet-base-v2}, 2021.

\bibitem{devlin2018bert}
J.~Devlin, M.-W. Chang, K.~Lee, and K.~Toutanova, ``Bert: Pre-training of deep bidirectional transformers for language understanding,'' \emph{arXiv preprint arXiv:1810.04805}, 2018.

\bibitem{hsu2021hubert}
W.-N. Hsu, B.~Bolte, Y.-H.~H. Tsai, K.~Lakhotia, R.~Salakhutdinov, and A.~Mohamed, ``Hubert: Self-supervised speech representation learning by masked prediction of hidden units,'' \emph{IEEE/ACM Transactions on Audio, Speech, and Language Processing}, vol.~29, pp. 3451--3460, 2021.

\bibitem{scikit-learn}
F.~Pedregosa, G.~Varoquaux, A.~Gramfort, V.~Michel, B.~Thirion, O.~Grisel, M.~Blondel, P.~Prettenhofer, R.~Weiss, V.~Dubourg, J.~Vanderplas, A.~Passos, D.~Cournapeau, M.~Brucher, M.~Perrot, and E.~Duchesnay, ``Scikit-learn: Machine learning in {P}ython,'' \emph{Journal of Machine Learning Research}, vol.~12, pp. 2825--2830, 2011.

\bibitem{breiman2001random}
L.~Breiman, ``Random forests,'' \emph{Machine learning}, vol.~45, pp. 5--32, 2001.

\bibitem{despotovic2016evaluation}
M.~Despotovic, V.~Nedic, D.~Despotovic, and S.~Cvetanovic, ``Evaluation of empirical models for predicting monthly mean horizontal diffuse solar radiation,'' \emph{Renewable and Sustainable Energy Reviews}, vol.~56, pp. 246--260, 2016.

\bibitem{hernandez2000application}
J.~Hernandez-Arteseros, R.~Compano, R.~Ferrer, and M.~Prat, ``Application of principal component regression to luminescence data for the screening of ciprofloxacin and enrofloxacin in animal tissues,'' \emph{Analyst}, vol. 125, no.~6, pp. 1155--1158, 2000.

\bibitem{brownless2020nested}
\BIBentryALTinterwordspacing
{Jason Brownlee}. {"Nested cross-validation for machine learning with python"}. 2020. [Online]. Available: \url{https://machinelearningmastery.com/nested-cross-validation-for-machine-learning-with-python/}
\BIBentrySTDinterwordspacing

\end{thebibliography}

\end{document}